\documentclass[notitlepage,11pt]{article}
\usepackage{fullpage}
\usepackage{subfigure}
\usepackage{latexsym}
\usepackage{graphicx}
\usepackage[intlimits]{amsmath}
\usepackage{booktabs}
\usepackage{amsthm,amstext,amssymb}
\usepackage[authoryear,round]{natbib}
\usepackage{amsfonts}
\usepackage{url}
\usepackage{ulem}
\usepackage{multicol}

\usepackage{setspace}
\def \s.t. {{\rm \;subject \;to \;}}

\def\bfd {{\bf d}} 
                                                                   
\def\bfb {{\bf b}}

\def\bfLambda {{\bf \Lambda}}

\def\bfy {{\bf y}}                                                                   
\def\bfY {{\bf Y}}

\def\bfp {{\bf p}}

\def\bfOmega {{\bf \Omega}}

\def\bfpi {{\boldsymbol\pi}}
\def\bftau {{\boldsymbol\tau}}

\begin{document}

\begin{center}
  {\LARGE Synonymous and Nonsynonymous Distances Help Untangle Convergent Evolution and Recombination}\\\ \\
  {
  Peter B.~Chi$^{1}$,
  Sujay Chattopadhyay$^{2}$,
  Philippe Lemey$^3$,
  Evgeni V. Sokurenko$^{2}$ and
  Vladimir N.~Minin$^{3,*}$\\   
    $^1$Department of Statistics, California Polytechnic State University, San Luis Obispo, CA, 93407, USA\\
    $^2$Department of Microbiology, University of Washington, Seattle, WA, 98195, USA\\
    $^3$Department of Microbiology and Immunology, Rega Institute, KU Leuven - University of Leuven, B-3000 Leuven, Belgium\\
    $^3$Departments of Statistics and Biology, University of Washington, Seattle, WA, 98195, USA\\
    $^*$Address for correspondene: \url{vminin@uw.edu}
  }
\end{center}

\begin{abstract}
When estimating a phylogeny from a multiple sequence alignment, researchers often assume
the absence of recombination. However, if recombination is present,
then tree estimation and all downstream analyses will be impacted, because different segments of the sequence
alignment support different phylogenies. Similarly, convergent selective pressures at the molecular level can 
also lead to phylogenetic tree incongruence across the sequence alignment. Current
methods for detection of phylogenetic incongruence are not equipped to distinguish between these two different mechanisms and assume that the incongruence is a result of recombination or other horizontal transfer
of genetic information. We propose a new recombination detection method that can make
this distinction, based on synonymous codon substitution distances.
Although some power is lost by discarding the information contained in the nonsynonymous substitutions, 
our new method has lower false positive probabilities than the comparable recombination detection method when the phylogenetic incongruence signal is due to convergent evolution. 
We apply our method to three empirical examples, where we analyze: 1) sequences from a transmission network of
the human immunodeficiency virus, 2) \textit{tlpB} gene sequences from a geographically diverse
set of 38 \textit{Helicobacter pylori} strains, and 3) Hepatitis C virus sequences sampled longitudinally from one patient.
\end{abstract}

\newpage

\section{Introduction}
The field of phylogenetics aims to describe evolutionary relationships among homologous sequences, by inferring a phylogeny, or evolutionary tree \citep{Felsenstein_2004}. 
In the estimation of a phylogenetic tree from a molecular sequence alignment,
the absence of recombination is frequently assumed, meaning that every site
along the sequence alignment has the same evolutionary history/phylogeny. Implications of 
recombination on tree estimation \citep{Posada_2002, Felsenstein_2004} and downstream
analyses \citep{Schierup_2000, Anisimova_2003} have motivated the development of a plethora of tests for the presence of recombination 
\citep{Awadalla_2003, Martin_2011}. 
Most of these methods try to test whether there are segments of the sequence alignment that support different phylogenies; if so, such phylogenetic incongruence is used as evidence of recombination \citep{Grassly_1997, McGuire_1997, Posada_2001}. However, another evolutionary force can produce an observed data pattern similar to the one produced by recombination. Suppose that the same selective pressure acts upon two sequences to make them appear more closely related to each other than they are under the true evolutionary history. Now, if this phenomenon, known as convergent evolution \citep{Wake_2011}, occurs between these two sequences only at a localized region of the alignment, then it will appear as if this region has a different evolutionary history than the remainder of the alignment, leading to an observed phylogenetic incongruency. To our knowledge, no existing method for
detecting phylogenetic incongruence can distinguish between recombination and convergent evolution. In this paper, we develop one method that can accomplish this task.
\par As a starting point, we consider the Dss method proposed by \citet{McGuire_1997} and 
implemented in the TOPALi software \citep{Milne_2004}. Dss, an
abbreviation for ``difference in the sum of squares,'' is a sliding window
approach that scans across the sequence alignment in question, with the assumption
that if a recombination breakpoint is present within any given window, then the portions of the window
on opposite sides of the breakpoint would have distinct evolutionary trees. The
test statistic produced by the Dss method is based upon the pairwise distance matrices for each half of every
window across the alignment, and an extreme value of the statistic is expected to
occur in a window that contains a recombination event at its center.
Our proposed modification is to base the test statistic on a measure of evolutionary distance that considers
only synonymous substitutions: the codon changes that do not result in amino acid changes. 
Assuming that selection predominantly acts on the amino acid level, it follows that selection has mainly 
an effect on nonsynonymous substitutions. 

Since synonymous substitutions provide `neutral' information about evolutionary relationships
of sequences under study \citep{Lemey_2005, Yang_2006, Obrien_2009}, we postulate
that using a distance metric which considers only synonymous 
substitutions within the Dss framework would still allow
for recombination detection, but will avoid the false positives 
resulting from convergent evolution. We develop a new statistic, and a novel parametric
bootstrap method to access the distribution of this statistic under the null hypothesis
of no recombination.
\par To test our new recombination detection method, we first proceed via simulations to 
compare its performance to the original Dss statistic, both in terms of their ability to identify
true recombination events, and to avoid false positives due to convergent evolution. 
We also examine three real data examples. The first is a human immunodeficiency virus (HIV) dataset, which
comes from nine Belgian patients that belong to a known HIV transmission chain \citep{Lemey_2005}. 
This dataset has been of particular interest because phylogenetic reconstructions can be compared to the known transmission chain, 
providing a real data example in which estimation procedures can be validated. In
their work, \citet{Lemey_2005} studied two distinct HIV genes: \textit{pol} and \textit{env}, and concluded that
the \textit{pol} gene was under convergent selective pressures, whereas the \textit{env} gene was not. Here, we
revisit this question with our method, by examining a concatenated alignment of the
\textit{pol} and \textit{env} genes. 
Our second real data example is a sequence alignment of the \textit{tlpB} gene encoding the methyl-accepting 
chemotaxis protein in \textit{Helicobacter pylori}. The importance of the TlpB protein lies in its role as a 
chemoreceptor and also in colonizing the bug to the gastric mucosa of its host \citep{Croxen_2006}. 
Interestingly, using multiple recombination detection statistics, we found evidence of recombination in 
\textit{tlpB}. Therefore, we choose this important gene to analyze the distribution of recombination signals across 
synonymous and nonsynonymous changes via our Dss statistics, and to determine evidence of actual recombination events. 
Finally, we investigate a Hepatitis C virus (HCV) sequence alignment from serum samples collected over roughly 10
years from one chronically infected individual \citep{Palmer_2012}. In their work, \citet{Palmer_2012} 
examined sequences of the hypervariable region 1 (HVR1) of the HCV genome and found
evidence of recombination between two distinct viral populations residing in the individual. 
However, HVR1 is subject to selective pressure, as antibody responses to HCV infection target this region \citep{Zibert_1995}. 
Thus, we analyze
this dataset with our Dss statistics to again test whether the recombination signal is due to a
true recombination, or due to convergent evolution.

\section{Methods}
\subsection{Evolutionary Distances}
\par Let $\bfY$ be a matrix that
represents a DNA sequence alignment, composed of row vectors $\bfy_1, \ldots, \bfy_n$, where
$n$ is the number of taxa/species/sequences. 
Then, $\bfy_k = (y_{k1}, \ldots, y_{kL})$, where $L$ is the length, or number
of sites in the sequence alignment. For a given DNA sequence alignment, one 
common summary of the data is the distance matrix $\mathbf{d} = \{d_{kl}\}$, where
each element $d_{kl}$ is the distance between sequences $k$ and $l$, for $k,l \in (1, \ldots, n)$. 
Intuitively, each pairwise distance
simply indicates how different two sequences are from each other. For example, two sequences
that are identical at every site along the alignment would
have a distance of 0, under most sensible measures of distance.
\par Typically, one assumes a substitution model that is defined by the
rates of change between the possible states. 
For a pair of taxa, the evolution of each site ($y_{ks},y_{ls}$) for $s \in (1, \ldots, L)$ can thus be described by
a continuous-time Markov chain (CTMC) with infinitesimal generator $\bfLambda = \{\lambda_{ij}\}$
for $i,j \in (1, \ldots, M)$,
where $M$ is the number of states (e.g. for DNA nucleotide data, $M=4$ as the
state space is $\{A,C,G,T\}$; for
DNA codons, $M=64$ as the state space
is $\{A,C,G,T\}^3$), and the rate of leaving state $i$ is $\lambda_i \equiv \sum_{j \neq i}^M \lambda_{ij}$.
With stationary
distribution $\bfpi \equiv (\pi_1, \ldots, \pi_M)$, we then can calculate distances for each pair $(k,l)$ as
\begin{eqnarray}
\label{eqn:edist}
\hat{d}_{kl} = \sum_{i=1}^M \hat{\pi}_i \hat{\lambda}_i,
\end{eqnarray} 
with the necessary 
parameter estimates being calculated from the data. As this quantity is equal to the average number of jumps
in a stationary continuous-time Markov chain, evolutionary distances are thus defined as the expected
number of substitutions per site, according to the given continuous-time Markov chain model. 

\par The notion of an evolutionary distance can be generalized to consider certain subsets of substitutions.
For example, it is sometimes of biological interest to count only 
transitions ($A \rightleftharpoons G$ and $C \rightleftharpoons T$), or
transversions ($A \rightleftharpoons T$, $A \rightleftharpoons C$, 
$G \rightleftharpoons T$, and $G \rightleftharpoons C$). A 
variety of \textit{ad-hoc} strategies could be used to account for this
\citep{Felsenstein_2004}, but it can also be formally incorporated into the framework of CTMC models of DNA evolution
as was demonstrated by \citet{Obrien_2009}. First, we define the set $\mathcal{L}$ to be the subset of
the lattice $\{1, \ldots, M\}^2$ that indicates the substitutions
which we wish to count; that is, $(i,j) \in \mathcal{L}$ if $i \rightarrow j$ is a substitution of interest. 
Then, we can express distances for any labeled subset of substitutions as
\begin{eqnarray}
\label{distlbl}
\hat{d}_{\mathcal{L}}=\sum_{i=1}^M \hat{\pi}_i \sum_{j\neq i}^M \hat{\lambda}_{ij} 1_{\{(i,j) \in \mathcal{L}\}},
\end{eqnarray} 
for each pair of sequences $(k,l)$.
The indicator function $1_{\{(i,j) \in \mathcal{L}\}}$ in (\ref{distlbl}) is equal to $1$ if $i \rightarrow j$ 
is a substitution of interest, and $0$ if it is not. In this manner, the labeled distance metric does not
count the substitutions that are not of interest. In this work, we specifically 
appeal to the labeled subset known as synonymous substitutions: the changes
in codon state which do not result in a change in amino acid.

\subsection{Distance-Based Recombination Detection: Dss Statistic}
\label{dss:def}
\par The Dss method for the detection
of recombination is based upon evolutionary distances. First, with an estimated distance matrix, 
one can 
infer a phylogeny, which consists formally
of a topology $\bftau$ and branch lengths $\bfb$. The tree distance $t_{kl}(\bftau, \bfb)$ between taxa $k$ and $l$ is 
then the sum of the branch lengths between them
on any particular topology. With $\hat{d}_{kl}$ estimated from DNA sequence data as above, 
least squares phylogenetic inference then proceeds by finding
\begin{eqnarray}
\label{lscritrec}
{\underset{\bftau, \bfb} {\text{argmin}}} \sum_{k,l} \left[\hat{d}_{kl} - t_{kl}(\bftau, \bfb)\right]^2,
\end{eqnarray}
which is the usual least squares criterion. 
The solution $(\hat{\bftau}, \hat{\bfb})$ to (\ref{lscritrec}) then gives the least
squares phylogeny.

\par
\begin{figure}[htb]
\begin{center}
\includegraphics[width=\textwidth]{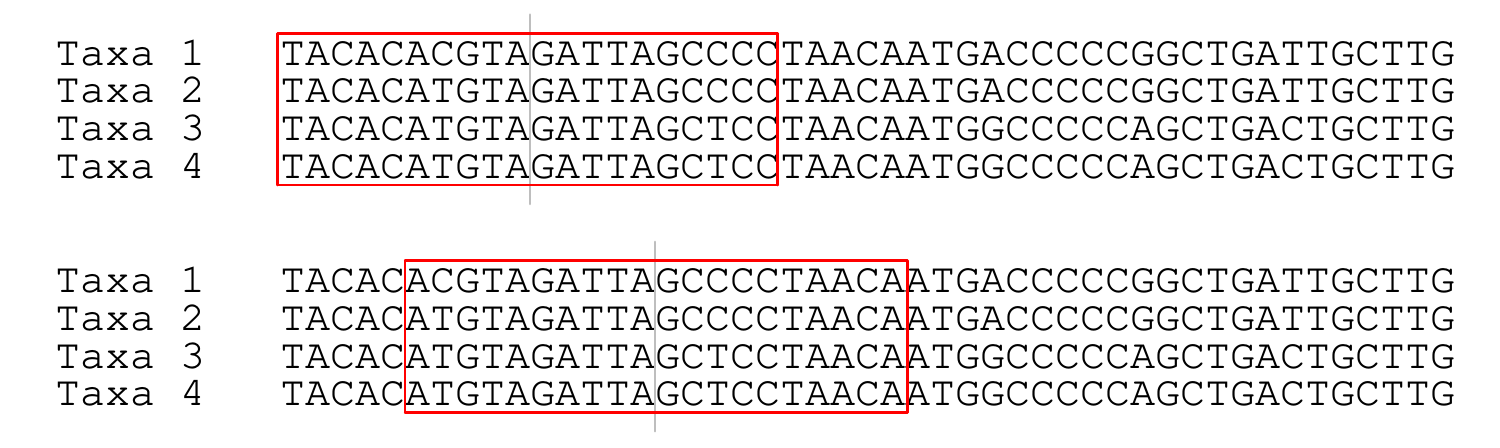}
\end{center}
\caption{Illustration of two overlapping sliding windows, shown as red boxes, across a sequence alignment of four taxa.
The vertical grey lines divide each window in half.}
\label{fig:window}
\end{figure}

 Now, we define 
recombination as an exchange of genetic material between two taxa that results in 
different evolutionary histories for the different respective parts of the sequence alignment. 
The Dss method then uses a sliding window approach \citep{McGuire_1997, McGuire_2000, Milne_2004},
as illustrated in Figure \ref{fig:window}, where the two panels show a window (in red) moving across
a sequence alignment. 
\par First, the average of all estimated pairwise distances from the entire sequence alignment
is recorded as $\overline{d}$. Next, the distance matrix is estimated with a
DNA substitution model for the first half
of a given window, along with its mean, $\overline{w}^{\{1\}}$. This distance matrix is then
standardized by multiplying each entry by $\overline{d}/\overline{w}^{\{1\}}$,
and the resulting standardized distance matrix for the first half of the window is recorded as $\hat{\bfd}^{\{1\}} = \{\hat{d}_{kl}^{\{1\}}\}$. Using
phylogenetic least squares as described above, we then calculate
\begin{eqnarray}
\label{ssaf}
{\underset{\bftau, \bfb} {\text{argmin}}}  \sum_{k,l} \left[\hat{d}_{kl}^{\{1\}} - t_{kl}(\bftau,\bfb)\right]^2.
\end{eqnarray}
The estimated topology is recorded as $\hat{\bftau}^{\{1\}}$, and the minimized value of the sum of squares in (\ref{ssaf})
is recorded as $SSa^F_w$. 
\par For the second half of the window, again the distance matrix is estimated, with its mean stored as $\overline{w}^{\{2\}}$
and again the standardized distance matrix is calculated by multiplying each entry by $\overline{d}/\overline{w}^{\{2\}}$ to
obtain $\hat{\bfd}^{\{2\}} = \{\hat{d}_{kl}^{\{2\}}\}$. Now, we calculate
\begin{eqnarray}
\label{ssbf}
SSb^F_w = {\underset{\bfb} {\text{min}}} \sum_{k,l} \left[\hat{d}_{kl}^{\{2\}} - t_{kl}(\hat{\bftau}^{\{1\}},\bfb)\right]^2.
\end{eqnarray}
That is, the topology from the first half is imposed as fixed in (\ref{ssbf}), and only the branch lengths are
optimized according to the sequence alignment of the second half of the window. 
\par For each window, we then have $Dss_w^F = (SSa_w^F - SSb_w^F)$. The entire procedure is repeated in the reverse direction,
by starting with a window at the end of the alignment, swapping the roles of each half of the window, and then sliding
it backwards across the sequence alignment; this gives $Dss_w^B$ for each window. Then, $Dss_w = \text{max}(Dss_w^F, Dss_w^B)$. 
Finally, here we consider only the maximum Dss statistic from all windows, giving us $Dss_{max} = {\underset{w} {\text{max}}}(Dss_w)$. 
\par Our modification of the Dss statistic uses estimates
of labeled distances for synonymous substitutions, estimated by the method developed by \citet{Obrien_2009}, 
replacing $\overline{d}$ and $\hat{d}_{kl}^{\{\cdot\}}$ in the calculation of $Dss_{max}$ above. 
\label{text:test1} The original 
Dss method tests for discrepant phylogenies throughout
windows across the sequence alignment. However, it cannot distinguish between the case where the 
discrepancies are actually due to an exchange of genetic material, as opposed to convergent selective pressure. 
In other words, with the Dss method, the null hypothesis is that the sequence alignment has one true evolutionary history that has
been affected neither by recombination nor convergent evolution, and evidence against the null hypothesis 
may be due to either recombination or convergent evolution, or due to the presence of both of these events. 
\par With our new synonymous Dss statistic, our null hypothesis remains the same, but our aim is
that our new test will detect only departures from the null that are due to an exchange of genetic material.
Under the assumption that selective
pressure acts on the amino acid level, synonymous substitutions are
presumed to be neutral, and therefore distances based upon them would ignore selective
pressures. In this manner, potential false positive signals for recombination due
to selection can be avoided. 

\subsection{Modified Parametric Bootstrap}
\par To assess statistical significance, \citet{McGuire_2000} propose a parametric bootstrap to generate
the null distribution of the test statistic.
Under this parametric bootstrap, the distribution of $Dss_{max}$ is simulated under the null
hypothesis as follows: first, the Ordinary Least Squares 
tree for the entire sequence alignment is obtained, under the
chosen model of substitution.
In this manner, the data are treated as if the sequences were inherited through one true tree and substitution model, in accordance
with the null hypothesis. Next, sequence data are
simulated under this tree, $B$ times. 
Finally, the Dss values are calculated under the same procedure
as outlined above for each simulated sequence alignment, and 
saving only the maximum from each simulated realization.
Thus, one obtains the distribution of the maximum Dss statistic
under the null hypothesis. This gives the basis for determining how extreme an observed
Dss statistic is, and we calculate the Monte Carlo estimate of the p-value as the proportion
of simulated null Dss values that are more extreme than our observed value in question. 
\par However, we must make important modifications to the parametric bootstrap for assessing
statistical significance of the Dss statistic as first proposed by \citet{McGuire_2000}. 
First, we estimate the distances between sequences on the codon scale (e.g. the expected number of substitutions
per codon site). Using this distance matrix, we then estimate the least squares
tree, which represents the null hypothesis of the evolutionary history of the sequences: with no recombination
or convergent evolution. Next, we estimate codon substitution parameters from the codon
model proposed by \citet{Nielsen_1998}: $\kappa$ (the transition/transversion ratio)
and $\bfOmega= (\Omega_1, \Omega_2, \Omega_3)$ where each $\Omega_i$ is
a nonsynonymous/synonymous rate ratio, with corresponding proportions $\bfp = (p_1, p_2, p_3)$
where each $p_i$ represents the probability that $\Omega_i$ will be selected as the nonsynonymous/synonymous
rate ratio for any given site. 
This mixture of three codon models allows for estimation of variable nonsynonymous/synonymous 
rate ratios at each site, to simulate the bootstrap data as similarly as possible to the evolutionary process
that created the original data.
With the null evolutionary history, 
$\kappa$, $\bfOmega$ and $\bfp$ estimated from the sequence data, we then simulate our parametric
bootstrap sequence alignment datasets. Using these, we calculate the original Dss statistic
and the synonymous Dss statistic in the manner described above, to then
obtain the distribution of the maximum, for each.

\subsection{Implementation}
\par All analyses have been performed using the {\tt{R}} package {\tt{synDss}}, in which we implemented the proposed methodology. The package contains
our implementation of the original Dss method, our synonymous Dss method, and modified parametric bootstrap.
The source code and binaries are available at \url{http://evolmod.r-forge.r-project.org/#synDss}.

\section{Results}
\subsection{Simulations: Power and Type I Error}
\label{results1}
\par To assess performance of each statistic, we simulate sequences under a codon model using the
software package PAML
\citep{Yang_2007}. Three basic scenarios are considered: 1) null; 2) true recombination event; 3) localized
convergent evolution. For each scenario, we consider a sequence
alignment with five taxa, and
we set $\kappa=2$ (transition/transversion ratio), and sample $\Omega$ (nonsynonymous/synonymous
rate ratio) from a discrete mixture model with values $\bfOmega=(0.1,0.8,3.2)$. We used
three sets of sampling probabilities:
$\bfp_1=(0.74, 0.24, 0.02)$, $\bfp_2=(0.85, 0.14, 0.01)$ and $\bfp_3=(0.99, 0.009, 0.001)$ 
 to produce average synonymous substitution proportions of $50\%$, $60\%$ and $75\%$ respectively.  
\par
\begin{figure}
\begin{center}
\includegraphics[width=\textwidth]{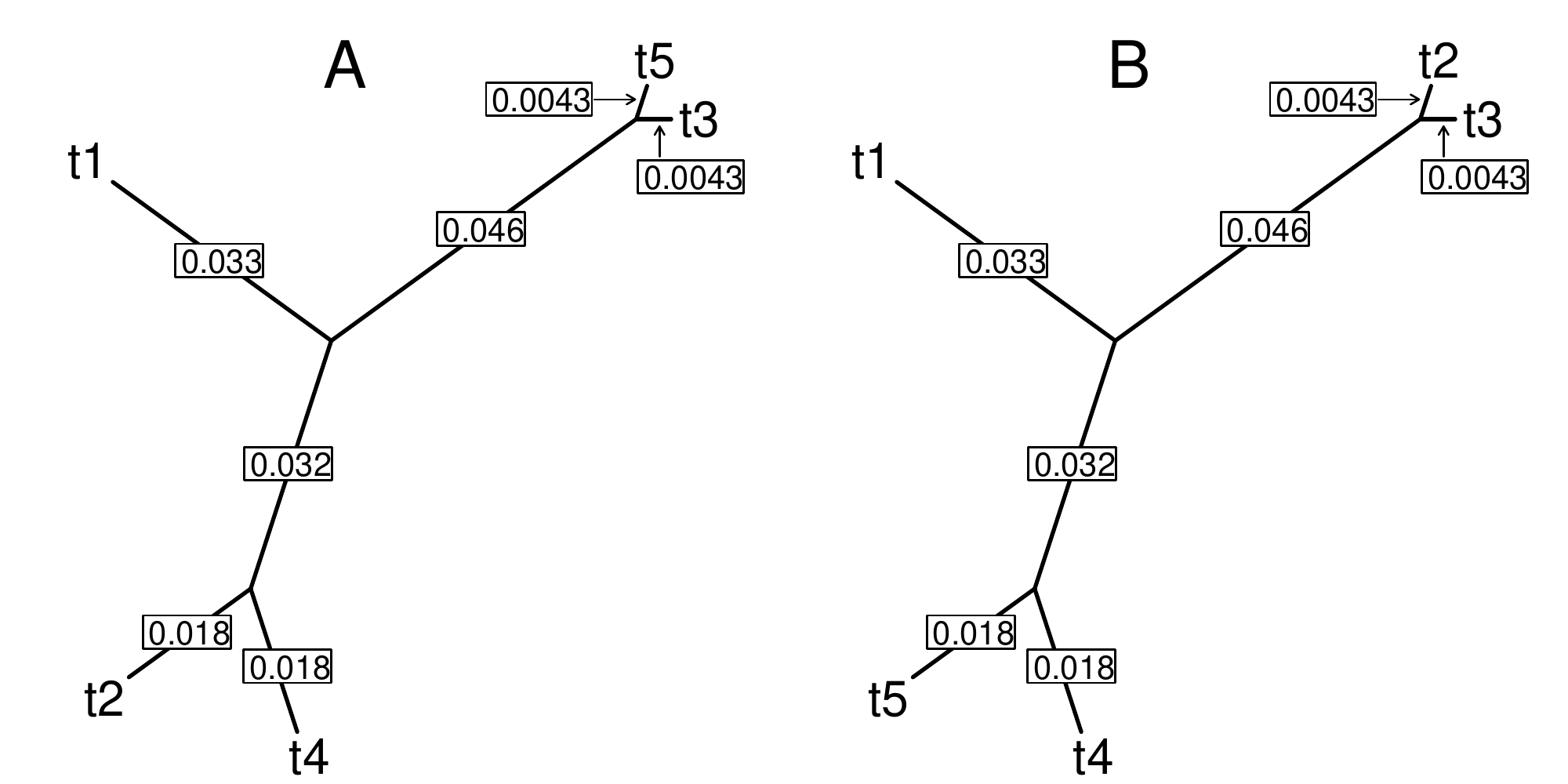}
\caption{Phylogenies used for simulations. Numbers indicate branch lengths, in expected number of substitutions
per site between two nodes.}
\label{fig:trees}
\end{center}
\end{figure}

 Under the null scenario, we assume that every site along the sequence alignment is inherited
according to one true evolutionary history. To simulate this, we provide PAML with one true phylogeny
with five tips (shown in panel A of Figure \ref{fig:trees}), 
and simulate codon sequences along this phylogeny. Each codon sequence consists
of 1032 codon sites (or 3096 nucleotides).
\par For the scenario with a true recombination event, we use two phylogenies 
corresponding to each side of the recombination
breakpoint (shown in panels A and B of Figure \ref{fig:trees}, which are
identical except that taxa 2 and taxa 5 are swapped). 
Partial sequence alignments of length 400 and 632 codons are simulated 
according to each phylogeny respectively, and then concatenated
to form one mosaic sequence alignment that is 1032 codons in length. 
\par For the scenario of localized convergent evolution,
we simulate codon sequences along one true phylogeny, and then choose a region upon which to have selection
act. Specifically, we use the latter 632 codons as this region (as in the recombination scenario). 
In this region, we again target taxa 2 and taxa 5, but here we make substitutions to change
differing amino acids each into another (concordant) amino acid. We choose only codon sites in which
this convergent evolution could occur with one nucleotide change in each of the sequences, and select
a proportion of these sites, uniformly at random, in which to make this change. To make
scenarios comparable, we convert the same proportion, on average, of all sites that have codon
variations initially: we set this proportion to $25\%$ in all cases. Noting that our convergent
evolution scheme can only act on sites which had nonsynonymous substitutions initially (since some sequences at these sites must be in different amino acid states), the proportion of eligible sites which get converted at random
must be adjusted according to the percentage of nonsynonymous substitutions in each
scenario (described above), in order to maintain the overall proportion of $25\%$ of all variable sites that will be converted.
\par Finally, for every scenario, we also vary the branch lengths, to effectively vary the number of substitutions, or
diversity, in each simulation. The branch lengths shown in Figure \ref{fig:trees} are the original set of branch lengths. 
We consider the original branch lengths, and also branch lengths that are scaled by $0.80$ and $0.67$ relative to the original
branch lengths, to result in scenarios of ``high diversity,'' ``medium diversity,'' and ``low diversity,'' respectively.

\par 
\label{text:typeI} With $\alpha=0.05$, Type I error probabilities under 1000 replicates of a simulated 
null scenario appear to be well-behaved, with estimated Type I error probabilities of $6.6\%$ and $6.3\%$
for the original Dss and synonymous Dss tests, respectively. Distributions of the p-values
resemble a uniform distribution, as shown in Supplementary Figure S1.

\par Next, we examine power to detect recombination, under the scenario with a true recombination event. We vary the expected number
of substitutions (diversity) and proportion of synonymous substitutions, and examine the corresponding effect on the power
of each version of the test statistic. Histograms of p-values from the original Dss
statistic and synonymous Dss statistic from one scenario are shown in Supplementary Figure S2, where
we observe $90\%$ power with the original Dss test statistic, and $76\%$ power with our synonymous Dss test statistic.
The results from all scenarios are shown in Table \ref{tab:recomb-pvals}. Our synonymous Dss statistic has reduced power
in every case, which is to be expected since we have reduced the amount of information used. The reduction
in power is less dramatic in the scenarios where a greater proportion of the substitutions are synonymous 
(bottom row of Table \ref{tab:recomb-pvals}), since less information is being discarded in these cases. 

\begin{table}[h]
\centering
\caption{Power of each test under the recombination scenario. Each column represents
one set of branch lengths (equivalently, the diversity), which correspond
to each average power of the original Dss test. Each cell represents the power of the synonymous
Dss statistic, with $95\%$ confidence intervals in parentheses. In each scenario, 100 
simulated replications were analyzed.}
\vspace{2mm}
\label{tab:recomb-pvals}
\begin{tabular}{cccc}
\cline{2-4}
& \multicolumn{3}{c}{Power of original Dss} \\
  \cline{2-4}
 & 99\% & 90\% & 85\% \\ 
  \hline
50\% syn &   66 (56.6, 75.4) &   38 (28.4, 47.6) &   20 (12.1, 27.9) \\ 
  60\% syn &   79 (70.9, 87.1) &   48 (38.1, 57.9) &   34 (24.6, 43.4) \\ 
  75\% syn &   87 (80.3, 93.7) &   76 (67.5, 84.5) &   62 (52.4, 71.6) \\ 
   \hline
\end{tabular}
\end{table}
\par
\begin{figure}
\begin{center}
\includegraphics[width=\textwidth]{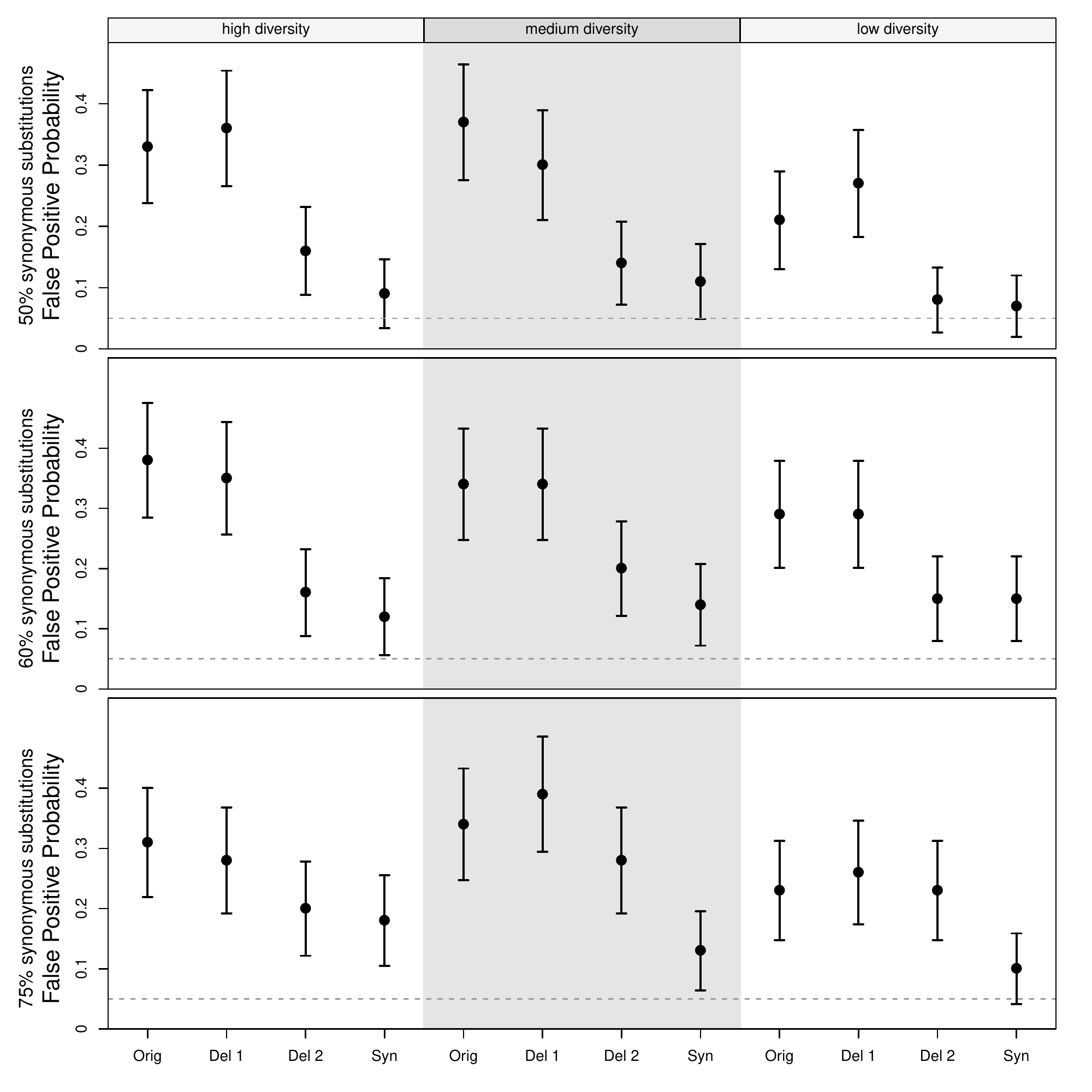}
\caption{False positive probability of each test under the convergent evolution scenario. Using the same
branch length sets (diversity) and synonymous substitution proportions as in the recombination scenarios, we induce
convergent evolution on the alignment instead of a true recombination. ``Orig'' refers to the original Dss statistic; ``Del 1''
refers to the case in which we remove a proportion of substitutions corresponding to the 
non-synonymous substitution proportion (and thus keeping a proportion corresponding to the synonymous substitution proportion);
``Del 2'' is similar to ``Del 1'' except that we also shrink the window size by the corresponding proportion; ``Syn'' refers to the
synonymous Dss statistic. Error bars represent 95\% confidence intervals based on the asymptotic binomial variance,
using the observed false positive probability as $\hat{p}$ to obtain standard errors. In each scenario, 100 simulated replications were analyzed.}
\label{fig:evol}
\end{center}
\end{figure}

 Under the scenario of convergent evolution, we compare the false positive probabilities under the original
Dss statistic and the synonymous Dss statistic. \label{text:test2} That is, while the Dss statistic
detects phylogenetic incongruence from any cause, we want to determine if the synonymous Dss statistic can avoid
giving a significant p-value when the phylogenetic incongruence is due to convergent evolution.
Under every scenario, we observe that the false
positive probability of the synonymous Dss method is substantially lower than that of the original Dss statistic, as shown
in Figure \ref{fig:evol}. For
example, under high diversity and $50\%$ synonymous substitutions, the estimated false positive probability for the
original Dss statistic is $33\%$, vs. $9\%$ for the synonymous Dss statistic. Results from all scenarios
are shown in Figure \ref{fig:evol}, labeled as ``Orig'' and ``Syn'' respectively.
\par We next examine whether this reduction in false positive probability 
might simply be due to the fact that the synonymous Dss statistic uses less information; that is, it considers only synonymous
substitutions. To answer this question, we first examine the effect of removing a proportion of sites, corresponding
to the proportion of synonymous substitutions. For example, under the scenario with $75\%$ synonymous substitutions, we
retain $75\%$ of the alignment sites at random, and then obtain the original Dss statistic. We observe that the false positive
rate under these simulations are similar to that of the original Dss statistic, as shown in Figure \ref{fig:evol}, 
labeled as ``Del 1.'' 
\par However, this effort suffers from the fact that, while the sequence alignments
are shorter, our window size has remained the same, thus resulting in fewer windows across the alignments. 
In our exploration of the Dss statistic behavior, we 
have noticed trends between window count and Power / Type I error (not shown) 
indicating that the ``Del 1'' regime is probably anti-conservative.
Thus, we perform another validation
experiment in which we also shrink the window size by the corresponding proportion; that is, if we removed
$50\%$ of the sites, we also shrink the window size by $50\%$. This is shown
in Figure \ref{fig:evol} as ``Del 2.'' Based on our experimentation with the relationship between window
size and power (not shown), we believe this to be a conservative
effort, and yet in all nine scenarios, we still obtain higher Type I error probabilities under this regime than that
of the synonymous Dss statistic.
\par Finally, for the scenarios with $50\%$ synonymous substitutions, we perform one additional set of experiments. Noting
that in these scenarios, a nonsynonymous Dss statistic would have, on average, the same loss of information as our
synonymous Dss statistic, we thus create a nonsynonymous Dss statistic in an analogous manner to which we created
our synonymous Dss statistic, using labeled distances for nonsynonymous substitutions. We then run this nonsynonymous
Dss statistic on the same set of data for each of the $50\%$ synonymous substitution scenarios. In the high diversity
case, we obtain a false positive probability of $19\%$, which is substantially higher than the synonymous Dss statistic's false
positive probability of $7\%$. For medium and low diversity, we obtain false positive probabilities of $13\%$ and $7\%$ respectively,
which are identical to their respective estimated false positive probabilities from the synonymous Dss statistic. 

\par For our simulation studies, we set the number of bootstrap replicates to $B=100$. We are
able to use $B=500$ for the real data analyses, but it would have been prohibitively time consuming to do this for the simulations. 
Although the resulting accuracy of significance level thresholds may thus be of some concern, we found through
a brief examination that the value
of the 95\% significance level threshold does not move substantially with $B=100$ on different parametric
bootstrap runs with the same original datasets.

\subsection{Data Analysis I: Belgian HIV Transmission Chain}
\par Phylogenetic analyses of HIV sequences are useful in characterizing its transmission and spread, and these analyses are
particularly relevant to elucidating the development of HIV drug resistance \citep{Lemey_2005}. 
However, while the typically high mutation rates and
short generation times for HIV are conducive towards a phylogenetic reconstruction, phylogenetic inference can be confounded by the high
recombination rates, and selective pressures 
imposed by antiretroviral therapy and
the host immune system \citep{Rambaut_2004}. Our method is the first to address the important issue of
distinguishing between recombination and convergent evolution, and thus we apply it here.
\par Of particular interest are the \textit{pol} and \textit{env} genes of HIV-1, which are responsible for replication \citep{Hill_2005}
and cell 
entry 
\citep{Coffin_1997}, respectively. These two genes were studied through
a transmission chain of nine Belgian HIV-positive patients \citep{Lemey_2005}, in which it was found
that a phylogenetic reconstruction using the sequenced \textit{env} gene was compatible with the known
transmission history among these nine patients; on the other hand, the phylogenetic reconstruction using the \textit{pol} gene sequences
was not compatible with the transmission history. This raised the question of whether selective pressures might be the cause of this
incongruity.
\par Specifically, \citet{Lemey_2005} explored whether the selective pressure may have been due to antiretroviral drug therapies
applied to HIV-positive patients in the transmission chain. They
hypothesized that patients on similar antiretroviral drug treatments may invoke convergent evolution on
their HIV strains, due to the fact that their respective HIV viruses may develop the same drug resistance-associated
mutations. By examinating known drug resistance-associated mutations within the \textit{pol} gene, they found
this was in fact the case with two of their individuals: ``Patient A'' and ``Patient I.'' That is, these two individuals
shared specific amino acid substitutions that have been identified by the International AIDS Society as being
associated with clinical resistance to HIV antiretroviral drugs \citep{Johnson_2003}.
\par Following this observation, \citet{Lemey_2005} 
then constructed phylogenetic trees for the \textit{pol} gene based on synonymous distances and nonsynonymous distances separately,
using the Syn-SCAN software \citep{Gonzales_2002}.
The synonymous tree was compatible with the transmission history, while the nonsynonymous tree was not, and showed
Patient A's strains clustering with those of Patient I. For an illustration, see Figure 4 from the work by \citet{Lemey_2005}. Thus, they
concluded that the \textit{pol} gene was under convergent selective pressure.
Here, we revisit this question by examining the behavior of each Dss statistic on a concatenated \textit{pol-env} sequence alignment. That is,
if we join the two sequence alignments together as one, will either recombination detection method indicate
the presence of intergenic or intragenic recombination? 

\begin{figure}[h]
\begin{center}
\includegraphics[width=\textwidth]{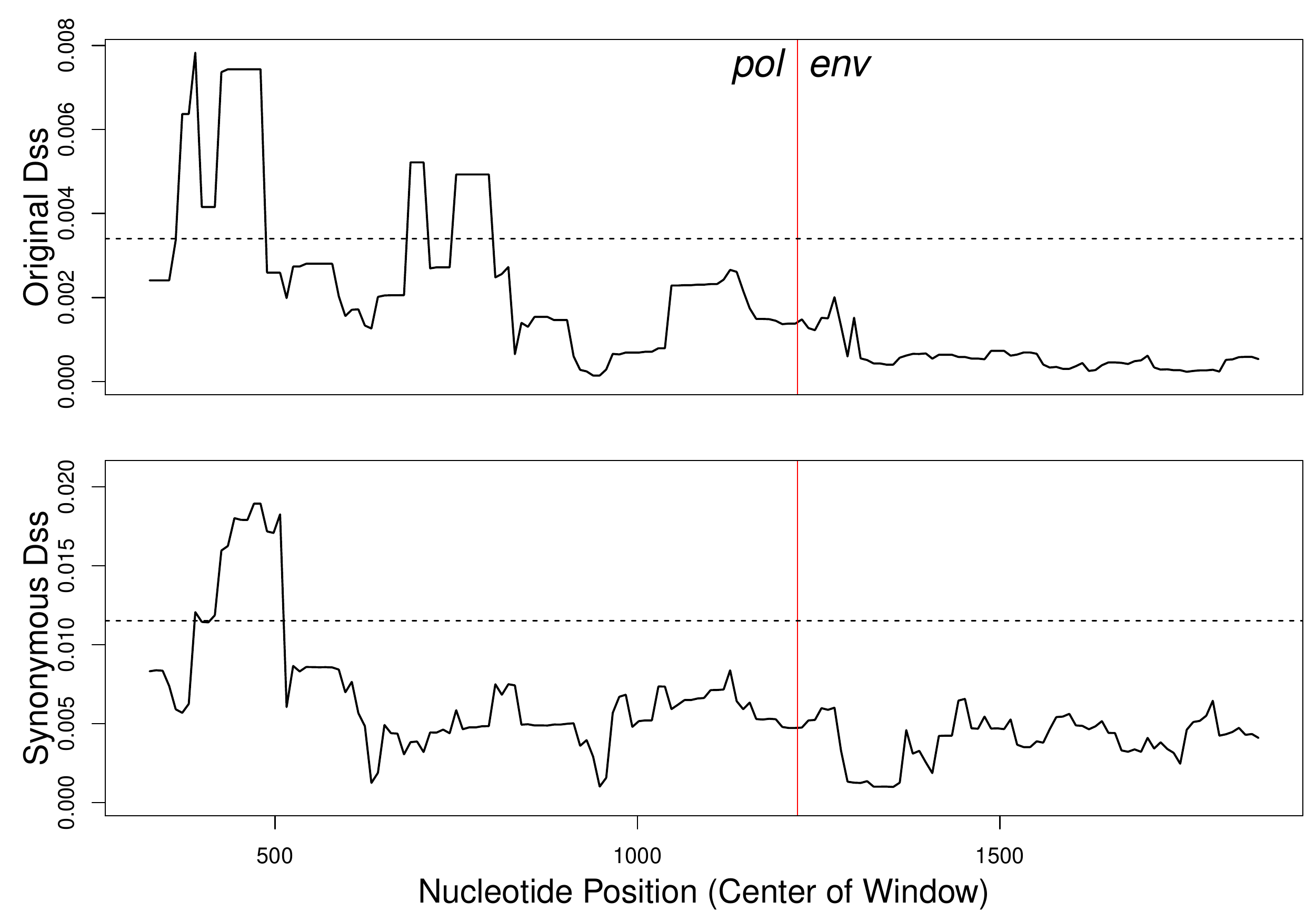}
\caption{\textbf{Dss statistic landscapes for \textit{pol-env} concatenation}. Dotted horizonal lines represent the $95\%$ significance
level for each test, from a parametric bootstrap with $B=500$. The red vertical lines represent the boundary between the two genes, with \textit{pol}
on the left, and \textit{env} on the right.}
\label{fig:polenv13}
\end{center}
\end{figure}
\par The dataset consists of nine individuals, with multiple samples taken longitudinally from some of them, for
a total of 13 sequences. Results from our analyses are shown in Figure \ref{fig:polenv13}. Our analysis used a window
size of 636 nucleotides (or 212 codons) and a step size of 9 nucleotides (or 3 codons). To assess statistical significance,
we used a parametric bootstrap with the number of replications set to $B=500$. We observe that both
the original Dss statistic and the synonymous Dss statistic cross their $95\%$ bootstrap significance thresholds, suggesting
the presence of a recombination event, as opposed to convergent evolution. 

\subsection{Data Analysis II: \textit{H. pylori} \textit{tlpB} gene}
\par One of the most common diseases in the world is chronic gastritis. 
The human-adapted motile Gram-negative bacteria \textit{Helicobacter pylori} is the major causative agent of 
chronic gastritis, in addition to causing stomach and duodenal ulcers and gastric cancer, thereby infecting about half 
of world's populations \citep{Feldman_1998}. 
Infection by \textit{H. pylori} is typically acquired
by ingestion, with person-to-person transmission occurring most commonly through vomit, saliva
or feces \citep{Feldman_2001, Parsonnet_1999}. Due to the emergence
of antibiotic-resistant strains, treatment of \textit{H. pylori} has begun
to fail in roughly 20--30\% of cases \citep{Graham_2009}, which points to the need for a better
understanding of the evolutionary processes that drive \textit{H. pylori} diversity
and survival.
\par Here, we focus on \textit{tlpB}, the gene that 
encodes the TlpB methyl-accepting chemotaxis protein. This protein
is crucial to \textit{H. pylori}'s ability to colonize
the stomach of its host, as it is responsible for its pH taxis, or movement in response
to high acidity \citep{Croxen_2006}. TlpB allows the bacterium to sense the pH of its surroundings,
and move toward an optimal pH zone. 
Due to these functional roles, TlpB is a potential target of the immune response by
the infected host. Thus, recombination is a potential diversification mechanism that could be used
by TlpB to avoid elimination by the host's immune response. To investigate this,
we selected \textit{tlpB} gene sequences from 38 completely sequenced \textit{H. pylori} genomes of globally representative isolates. 
Recombination detection analyses by PhiPack \citep{Bruen_2006} involving three different statistics -- 
Pairwise Homoplasy Index (p$<$0.0001), Maximum $\chi^2$ (p=0.005) and Neighbor Similarity Score (p$<$0.0001) -- 
detected evidence of recombination in this gene. 
\par However, there remained a possibility that this recombination signal was 
actually a result of convergent evolution. Especially for \textit{H. pylori} that shows extensive genomic diversity, it could be a 
common scenario that an excess of convergent mutations created an illusion of recombination, 
thereby confusing traditional recombination detection algorithms.
With the same \textit{tlpB} sequence alignment dataset from 38 isolates, our aim is to determine whether our Dss 
analysis concludes that there is truly a recombination signal, or whether it was actually
due to convergent evolution.

\begin{figure}[h*]
\begin{center}
\includegraphics[width=\textwidth]{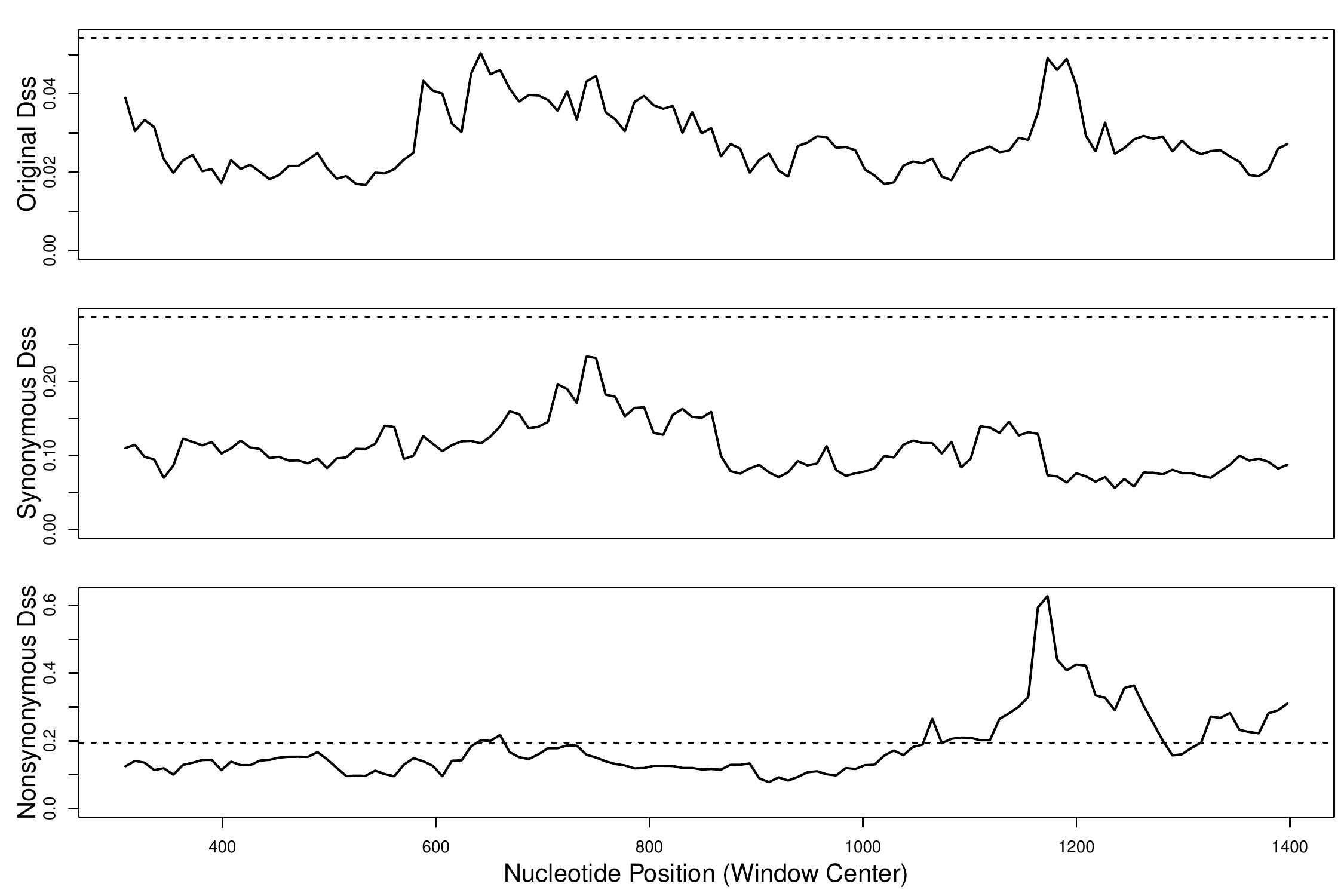}
\caption{Dss statistic landscape for \textit{H. pylori} \textit{tlpB} gene. 
Dotted horizonal line represents the $95\%$ significance level for each test, from a parametric bootstrap with $B=500$.}
\label{fig:tlpB}
\end{center}
\end{figure}\par If recombination has indeed occurred, then
we would expect that both the original Dss statistic and the synonymous Dss statistic would find
a statistically significant signal for recombination. In contrast, if the recombination signal is
actually due to convergent evolution, then
we would expect that the original Dss test and the nonsynonymous Dss test would find a statistically significant signal for recombination,
whereas our synonymous Dss test would not, as in the HIV example.
\par Here, we perform analyses with a window size of 600 nucleotides (200 codons), and step size of 9 nucleotides (3 codons).
The sequence alignment was 1695 nucleotides (565 codons), yielding approximately 122 windows across
the alignment. To obtain
significance threshold levels, we use the parametric bootstrap with the number of replications set at $B=500$.
Results from our analyses are shown in Figure \ref{fig:tlpB}. We observe that neither the original Dss
statistic nor the synonymous Dss statistic crosses its respective $95\%$ bootstrap significance threshold. However,
the original Dss statistic was very close, and in fact gave a bootstrap p-value of 0.058. In contrast,
the synonymous Dss statistic did not come quite as close to its $95\%$ bootstrap significance threshold,
and gave a bootstrap p-value of 0.14. 
\par It might be a possibility that the lack of signal with the synonymous Dss statistic was due to a loss of power. However, we found that 
the nonsynonymous Dss statistic did cross its
$95\%$ bootstrap significance threshold, with a bootstrap p-value of 0.
Also, an examination of the sequence alignment revealed
that there was approximately a 1.5:1 ratio of synonymous substitutions to nonsynonymous
 substitutions. Therefore, any loss of power observed with the synonymous Dss statistic should also
 be observed with the nonsynonymous Dss statistic. More importantly, these results demonstrated that the recombination 
 signal was driven primarily by
 nonsynonymous substitutions. Absence of any such signal with the synonymous Dss statistic strongly suggests that the
 recombination signal in the nonsynonymous changes was actually due to convergent evolution, most likely in response to adaptive selection pressures.

\subsection{Data Analysis III: HCV}
\par HCV is an RNA virus that is estimated to infect roughly 3\% of the human population worldwide, and is a leading cause of liver
disease and liver cancer \citep{WHO_2003}. Overall, treatment success has been limited, and thus it has been recognized
that a greater understanding of the virus' evolutionary behavior is crucial to effective prevention and treatment
of HCV infection \citep{Gray_2012}. Indeed, specifically
the matter of whether genetic recombination occurs in HCV has important implications regarding resistance
development against antiretroviral treatments used against the diseases that are caused by HCV infection \citep{Morel_2011}.
\par Similar to HIV, HCV mutates very rapidly within an infected host, which makes
treatment difficult, but also should reveal patterns that will lead to a greater understanding of the link
between the evolution of the virus and progression of disease \citep{Okamoto_1992, Smith_1997}. 
Curiously, however, there have been few reports of recombination occuring in HCV, despite the fact that
recombination can be an important diversification mechanism in positive sense RNA viruses \citep{Gonzalez_2011}. One potential reason
is that simultaneous infection by two or more HCV types might be rare \citep{Viazov_2000, Tscherne_2007}. Additionally,
it has been postulated that the viability of recombinant strains is poor \textit{in vivo} \citep{Prescott_1997}.
\par Here, we investigate a sequence alignment of the hypervariable region 1 (HVR1) of HCV, from serial
samples taken over 9.6 years from a single infected patient \citep{Palmer_2012}.
In the initial study, \citet{Palmer_2012} found evidence of recombination between two HVR1 subpopulations
within the patient. However, this genetic region is subject to strong selective pressures as 
the envelope glycoprotein is targeted by the host antibody responses \citep{vonHahn_2007}. Thus, we analyze this sequence
alignment with our Dss statistics, to determine whether our analysis corroborates the original findings of \citet{Palmer_2012},
or whether this recombination signal is confounded by convergent evolution.

\begin{figure}[h]
\begin{center}
\includegraphics[width=\textwidth]{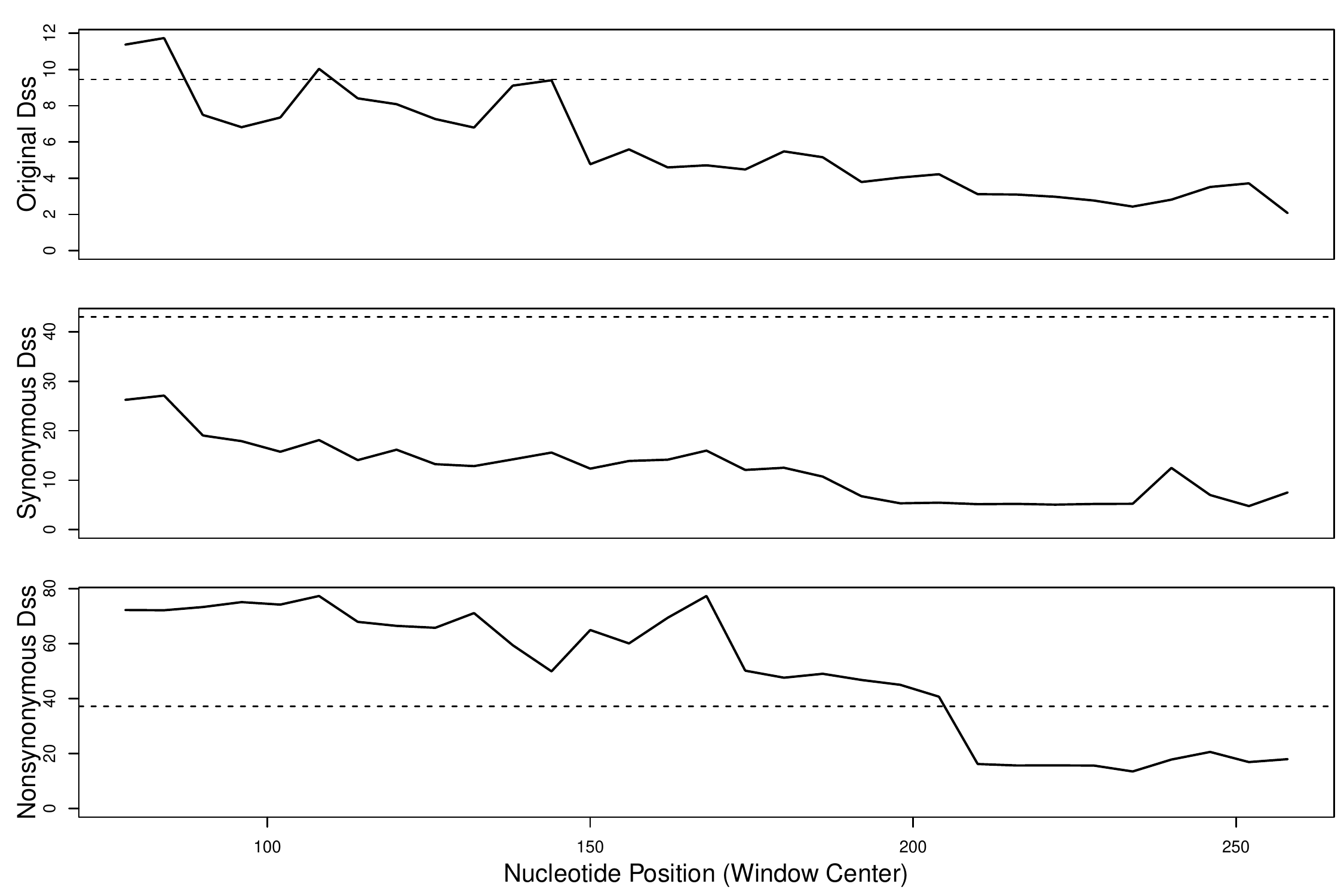}
\caption{Dss statistic landscapes for HVR1 of HCV alignment. Dotted horizonal lines represent the $95\%$ significance
level for each test, from a parametric bootstrap with $B=500$.}
\label{fig:hcv_500}
\end{center}
\end{figure}
\par We perform analyses with a window size of 144 nucleotides (48 codons) with a step size of 6 nucleotides (2 codons). The
sequence alignment was 324 nucleotides long (108 codons), which gives 31 windows across the alignment. Results
from our analyses are shown in Figure \ref{fig:hcv_500}, with significance threshold levels obtained from the parametric
bootstrap with $B=500$. We observe that the original Dss statistic crosses its $95\%$ bootstrap significance threshold,
whereas the synonymous Dss statistic does not. Additionally, the nonsynonymous Dss statistic crosses its
95\% bootstrap significance threshold, suggesting that the lack of signal with the synonymous Dss statistic was not
simply due to a loss of power. This is further supported by an examination of the raw counts
of synonymous vs. nonsynonymous substitutions in this sequence alignment, as there are in fact slightly more
synonymous substitutions than nonsynonymous substitutions.
Thus, we find statistically significant evidence that the recombination signal in HVR1 sequences
is actually due to convergent evolution.

\section{Discussion}
\par In this work, we have introduced the synonymous Dss statistic, developed to give a statistical method which
allows us to distinguish between recombination and convergent evolution. Our simulations show that while 
our synonymous Dss statistic loses some power compared to the original Dss statistic, it does have a lower
false positive probability when the signal is due to convergent evolution. Furthermore, we provide some verification that this lower
false positive probability is not simply due to the loss of power, as suggested by the false positive probabilities of
the various scenarios in which we remove a comparable portion of the information in the sequences, 
 shown in Figure \ref{fig:evol}. 
\par 
Our real data analyses highlight the usefulness of our methodology when dealing with situations where both recombination and
convergent evolutionary may be participating in the evolution of molecular sequences. 
We find evidence contrary to the conclusion by \citet{Lemey_2005}
that convergent evolution has occurred in the \textit{pol} gene lineage of HIV; instead,
we find evidence for the occurrence of a recombination event. The benefit of using our method is that
we have created a statistical testing framework for addressing precisely this question. In contrast,
\citet{Lemey_2005} examined phylogenies constructed with synonymous and nonsynonymous substitutions separately, basing their conclusion on whether
each phylogeny matched the known transmission chain. 
However, \citet{Lemey_2005} had difficulty providing a measure of statistical significance for their findings; 
in contrast, our method naturally assigns
statistical significance to our individual findings. Furthermore, \citet{Lemey_2005} implicitly assumed
that the entire \textit{pol} gene had one evolutionary history, and likewise for the \textit{env} gene. If recombination had occurred within either gene, 
then this assumption would be violated. Similarly, \citet{Lemey_2005} compared the synonymous 
and nonsynonymous trees of the entire \textit{pol} gene, while convergent evolution is most likely
to be localized to a subset of sites, making it difficult to detect using distances based on the whole alignment. 
In contrast, our sliding window method is able to separate the contributions of 
synonymous and nonsynonymous substitutions to the \textit{local} phylogenetic incongruence signal, 
which we believe to be an important advantage. 

\par \citet{Croxen_2006} found that \textit{tlpB}
 mutant strains were deficient in colonization due to their inability to respond to the pH
 gradient. More recently, engineered mutational analysis showed 
 the disruption in urea-binding and thermal stabilization of the mutational variants \citep{Sweeney_2012}. Also, 
 the work demonstrated reduced chemotactic responses of urea-binding variants to acid. These 
 experimental results suggested the possibility that the natural mutational variations in the TlpB protein 
 could arise from adaptive selection pressures. While the gene showed recombination signals via traditional 
 statistics, our novel approach detected the signal to be due to the presence of convergent nonsynonymous 
 (i.e. amino acid replacement) mutations. Such events of repeated, independent (i.e. phylogenetically unlinked) 
 accumulation of mutations at specific amino acid positions of encoded proteins represents powerful evidence of 
 adaptive events \citep{Christin_2012, Tenaillon_2012}. Taken together, our results, on one hand, 
 indicated the presence of adaptive evolution of the \textit{H. pylori tlpB} gene via convergent nonsynonymous mutations. 
 On the other hand, this study depicted the promise of our approach to differentiate convergent mutational events from recombination.
	\par As noted by \citet{Gonzalez_2011}, there has been some debate regarding the occurrence of recombination
	as a diversification mechanism in HCV. The reports of \textit{in vivo} recombination have been questioned
	as being due to either PCR artifacts or misidentification due to convergent evolution. Here, we find evidence that the
	recombination signal found by \citet{Palmer_2012} in their HVR1 sequence alignment
	is due to convergent evolution. As the occurrence of recombination in HCV continues to be called
	into question, our results side with the notion that empirical evidence of recombination of HCV sequences 
	should be interpreted with caution, because of a possibility of false positives due to convergent evolution.
\par 
A fundamental question that one might ask is how our method is advantageous over simply removing sites
that contain nonsynonymous substitutions during a recombination detection analysis. 
An illustration of the answer can be observed by considering an
alignment containing a large number of sequences, in which multiple substitutions per site would not be uncommon. 
Thus, if two substitutions had occurred at a particular site, then one
substitutions could be synonymous and the other could be nonsynonymous. To use the brute-force approach of removing
sites that contain nonsynonymous substitutions would necessarily remove the information contained
in the synonymous substitution that had occurred at that site; that is, to remove the site means to remove the entire
column from the sequence alignment, so all of the information contained in that site is lost. In contrast, our approach
of counting synonymous substitutions under the framework laid out by \citet{Obrien_2009} removes the nonsynonymous mutation
information in a more elegant manner, avoiding the total loss of information that would result from removing
entire sites.
\par A potential future development would be to create a coherent method to disintangle recombination and
convergent evolution without a convoluted three-way comparison, between the original Dss statistic, the synonymous Dss statistic,
and the nonsynoymous Dss statistic. That is, in this study, we would conclude that there is evidence for recombination
if both the original Dss statistic and the synonymous Dss statistic show a positive signal. If the original Dss statistic
shows a positive signal but the synonymous Dss statistic does not, then we would conclude that this is evidence of
convergent evolution, further validated if the nonsynonymous Dss statistic also showed a positive signal. It would
be preferable if a methodology could produce one coherent statistic to evaluate in order to answer
this question, instead of two or three.
\par Finally, there is the potential that our concept could be implemented in other recombination detection
regimes, specifically those that are likelihood-based. It is well documented that sliding window recombination 
detection methodology, such as that of the Dss statistic, has drawbacks. For example, the behavior of the test statistic is somewhat influenced by the window size chosen, and there are few guidelines on how to select this tuning parameter \citep{McGuire_2000}. 
Also, a multiple comparisons issue exists, since each window produces a value of the test statistic. Although this issue is
handled by considering only the maximum statistic value from the alignment and performing an appropriate 
parametric bootstrap test for statistical significance, this strategy prevents estimating locations of recombination break-points with confidence. Thus, it may be advantageous
to import our concept of synonymous recombination detection into a likelihood-based framework, such
as those proposed in \citep{Husmeier_2003}, or in \citep{Minin_2005,Suchard_2003}. 

\section*{Acknowledgements}
We thank Adam Leach\'{e} and Ken Rice for helpful comments and discussions. VNM was supported by the National Science Foundation grant DMS-0856099 and the National Institutes
of Health grant R01-AI107034. VNM and EVS were supported by the 
NIH ARRA award 1RC4AI092828. PL acknowledges funding from the European Research Council under the European Community's Seventh Framework Programme (FP7/2007-2013) under ERC Grant agreement no. 260864.

\bibliography{pc}{}
\bibliographystyle{plainnat}

\clearpage

\renewcommand{\thefigure}{S-\arabic{figure}}
\setcounter{figure}{0}

\begin{center}
{\Large Supplementary Figures}
\end{center}

\begin{figure}[h]
\includegraphics[width=\textwidth]{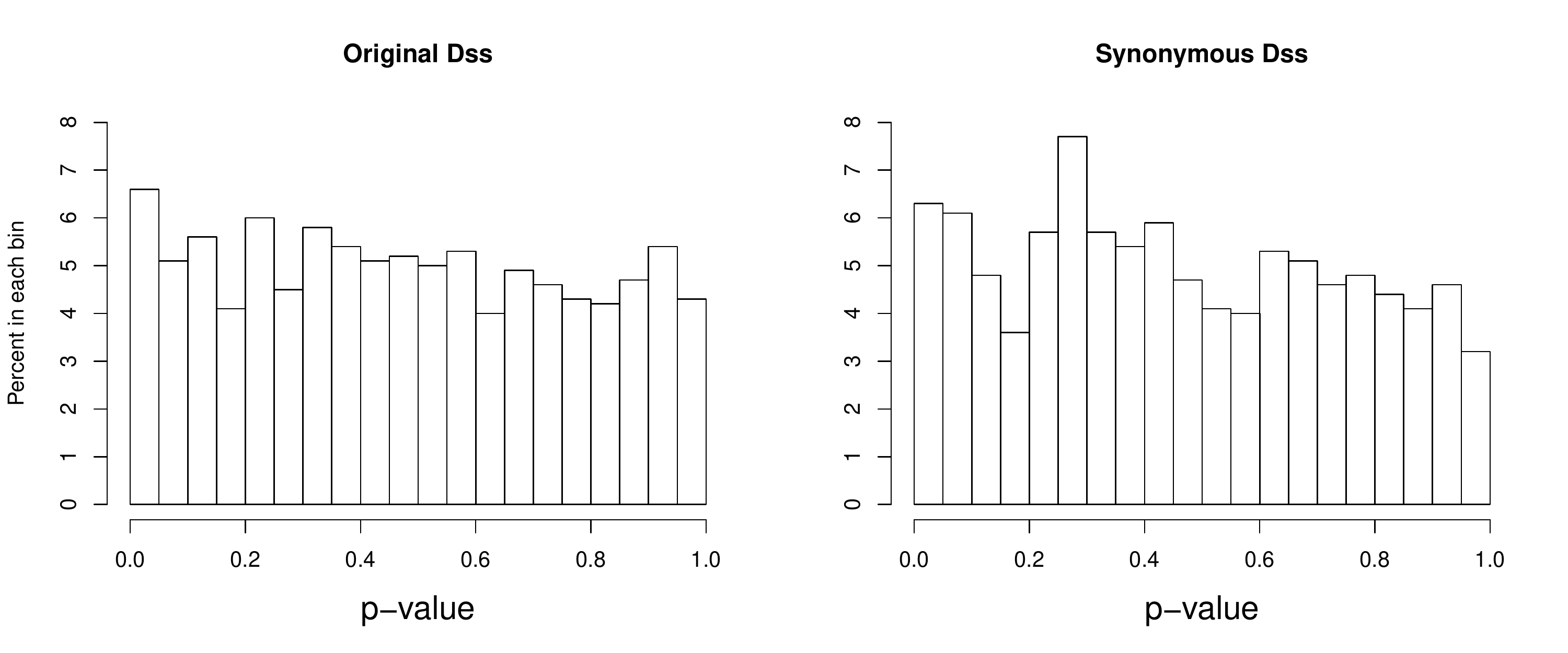}
\caption{Distribution of p-values of the original and synonymous Dss tests when the data are 
simulated under the null hypothesis of no recombination and no convergent evolution.}
\end{figure}

\vspace{1.5cm}

\begin{figure}[h]
\includegraphics[width=\textwidth]{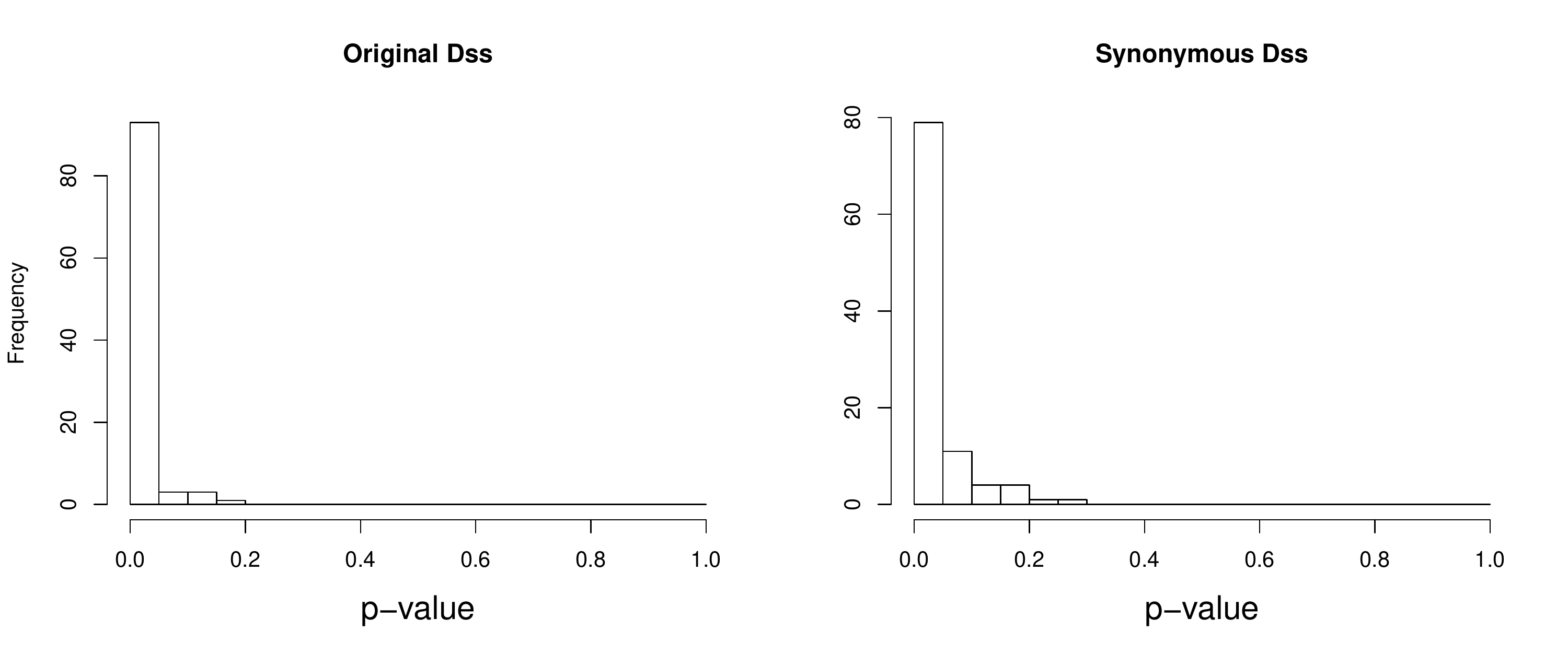}
\caption{Distribution of p-values of the original and synonymous Dss tests when the data are 
simulated under a model with recombination.}
\end{figure}

\end{document}